\documentclass[12pt,letterpaper]{article}
\usepackage[margin=1.2in]{geometry}
\usepackage{setspace}

\usepackage{epsfig, amsthm, amsmath, amssymb, mathrsfs, latexsym, xpatch, xcolor, mathtools, verbatim, hyperref}
\usepackage[mathscr]{euscript}
\usepackage[titletoc]{appendix}
\usepackage[normalem]{ulem} % strike out % for better underline, use \uline{...}
\usepackage{tabularx}
\usepackage{stmaryrd} % for linguistics symbols
\usepackage[margin=1cm]{caption}
\usepackage{float} % to make figure exactly here

\definecolor{cardinal}{rgb}{0.8, 0.0, 0.0}

\usepackage{mdframed}
\newcommand{\oop}{
	\begin{mdframed}[
	linecolor=lightgray, 
	backgroundcolor=gray!5,%white, 
	linewidth=1pt,
	innerleftmargin=15pt, innerrightmargin=15pt, innertopmargin=10pt, innerbottommargin=10pt
]}
\newcommand{\eed}{
	\end{mdframed}
}

\newcommand{\op}{\begin{itemize}}
\newcommand{\ed}{\end{itemize}}

\newcommand{\ope}{\begin{enumerate}}
\newcommand{\ede}{\end{enumerate}}

\newcommand{\xm}{\item[]}
\newcommand{\im}{\item}

\newcommand{\PP}{\mathbb{P}}

\title{Frequentist Statistics as Internalist Reliabilism}
\author{Hanti Lin \\[0.5em] University of California, Davis \\ika@ucdavis.edu}

\date{To Appear in \\[0em] Shan, Y. (ed.) {\em Integrating Philosophy of Science and Epistemology}, \\[0.3em] Synthese Library, Springer.}
\begin{document}

\maketitle

\begin{abstract} \noindent There has long been an impression that, for theories of justified belief or inference, reliabilism implies externalism and that frequentist statistics, due to its reliabilist nature, is inherently externalist. I argue, however, that frequentist statistics can plausibly be understood as a form of internalist reliabilism---internalist in the conventional sense, yet reliabilist in certain unconventional and intriguing ways. Crucially, in developing the thesis that reliabilism does not imply externalism, my aim is not to stretch the meaning of `reliabilism' merely to sever the implication. Instead, it is to gain a deeper understanding of frequentist statistics, which stands as one of the most sustained attempts by scientists to develop an epistemology for their own use.\end{abstract}

%\newpage
%\addcontentsline{toc}{section}{Table of Contents}
%\tableofcontents
%\newpage

\section{Introduction}

The internalism-externalism divide was originally formulated as a debate about the justification of belief, rather than inference. According to internalism, the factors determining whether an agent's belief is justified must, in some sense, be {\em internal} to that agent. Paradigm examples of internal factors include the following: (i) the beliefs one holds, (ii) the background assumptions one takes for granted, (iii) the propositions one adduces as reasons for some beliefs, and (iv) the deductive relations among the propositions involved in these attitudes---namely, believing, taking for granted, and adducing as a reason. Although drawing a precise boundary around internal factors is challenging, internalists generally emphasize that one's belief is justified exactly when it is possible, in principle, to articulate, {\em from within} one's first-person perspective, a justification for holding that belief (BonJour 2005). This ``from within'' perspective requires the factors of justification to be, in a sense, internal to the agent. Externalists, however, disagree. They argue that justification is not subject to such stringent requirements and that at least one factor of justification is, in a corresponding sense, external to the agent. A paradigm example of an external factor is the actual reliability of one's belief-producing process---its reliability in producing true beliefs (Goodman 1972).

% Or, in short, an agent's belief is justified exactly when it is possible for that agent to justify that belief. 
% Justifying is fundamentally a person's justifying. 
% What justifies your belief? Can you justify your belief?

Although the disagreement between internalists and externalists primarily concerns the justification of belief, it naturally extends to other areas. The objects of evaluation can include beliefs, acts, or inference methods. For instance, one might ask whether an agent's adoption of a particular inference method is justified or whether an inference method is justified within the agent's context of inquiry. Such questions are common in philosophy of science, where Hume's problem is often framed as a question about the possibility of justifying induction (Salmon 2017). Here, the focus will be on the justification of inference methods.\footnote
	{While the objects of evaluation can vary, the evaluative concept in question can also differ. It need not be being {\em justified}; it could, for instance, be being {\em appropriate}, {\em reasonable}, {\em good}, or {\em best}. However, I will focus on the concept of being justified, even though the points made below apply equally well to these other evaluative concepts.}

I will not address whether internalism or externalism is correct, nor will I delve deeply into which versions of these views are more plausible. My goal here is modest: to develop an intriguing example of an internalist theory. Let me explain.

I will argue that frequentist statistics is, or can plausibly be interpreted as, a form of {\em internalist reliabilism} about the justification of inference methods---despite the somewhat tacit but widespread belief that reliabilism implies externalism.\footnote{See Steup (2004) for a rare explicit exception.} The reason is straightforward: strictly speaking, reliabilism does not necessarily lead to externalism. To be sure, externalism does follow from the {\em conventional} version of reliabilism, which holds that justification depends on the {\em actual} reliability of the inference method or belief-producing process in question (Goodman 1972). However, if I am right, frequentist statistics can be interpreted as an unconventional version of reliabilism: whether an inference method $M$ is justified depends not necessarily on the actual reliability of $M$, but rather on its reliability in each possible world within a certain range---namely, those compatible with the background assumptions taken for granted in one's context of inquiry. Such an unconventional version of reliabilism does not conflict with internalism, as will be made clear below.

As a warm-up, I will begin by briefly explaining why frequentist statistics has long been regarded as largely externalist (Section \ref{sec-review}). Upon closer examination, however, frequentist statistics will be shown to allow for a natural interpretation as internalist in character (Sections \ref{sec-formal-def}-\ref{sec-diagnosis}). I will then take a further step to demonstrate that frequentist statistics can be understood as both internalist and reliabilist, with reliabilism manifesting in two unconventional but important senses that will be made precise below (Section \ref{sec-relia}). For simplicity, examples will primarily be drawn from one area of frequentist statistics: hypothesis testing. Yet I will briefly explain how the main idea extends to other areas, such as point estimation (Section \ref{sec-extension}).

A clarification before we begin: Throughout this paper, by `frequentism,' I mean a certain {\em epistemological} view about statistical inference. Despite its name, which has become too entrenched to change, the frequentist view does not necessarily involve frequencies. Stated more generally, frequentism holds that inference methods should be evaluated based on their reliability or unreliability, which, in turn, can be defined by physical objective probabilities of error. But what are physical objective probabilities---frequencies or something else? This is a {\em metaphysical} issue that the epistemological view leaves open. These probabilities might be best interpreted as frequencies (Neyman 1955), propensities (Popper 1959), or primitive physical states posited in science (Sober 2000: sec. 3.2). For clarity and focus, I will set aside the metaphysical debate and concentrate on epistemological issues. However, see Lin (2024$a$) for a substantive argument on why a clear separation between metaphysical and epistemological issues is necessary in philosophy of statistics.\footnote{Lin (2024$a$) argues that such a separation is crucial for categorizing some important philosophical views that have emerged in the history of statistics---a two-dimensional spectrum extending from radical frequentism to radical Bayesianism, with intermediate positions in between. One dimension corresponds to the metaphysical issue; the other, to the epistemological.}

%Data science, broadly construed to encompass statistics and machine learning, represents scientists' attempt to develop an epistemology for their own use. While there are several approaches to data science, frequentist statistics remains to be the dominant  statistics has long been thought to be {\em reliabilist}, in the sense of taking the reliability of an inference method or belief-producing procedure as one of the factors that determine whether the objects of evaluation are justified. But, notably, reliability is a classic example of an external factor, so reliabilism appears to imply externalism. Thus, So frequentist statistics appears to be externalist owing to its reliabilist nature. 

\section{Frequent Statistics: Externalist or Internalist?}\label{sec-review}

Suppose that a scientist is testing a hypothesis $H_0$ with a prescribed sample size $n$. An inference method for this task, or a {\em test}, is a function that outputs a verdict---either `\textsf{Reject $H_0$}' or `\textsf{Don't}'---when it receives a data sequence of the given length $n$. Tests are evaluated against certain standards, with the minimum qualification being a {\em low significance level}. This norm can be stated as follows:
	\oop 
	An inference method $M$ for hypothesis testing is justified only if it has a low significance level of $\alpha$ (say 5\%).
	\eed 
But how are significance levels defined? Introductory texts often define them somewhat informally, as follows (Hacking 1965/2016, p. 84; Howson \& Urbach 2006, p. 146; Rosner 2016, pp. 213-214):
	\oop 
	{\bf Informal Definition 1.} A test $T$ is said to have a low significance level (at level $\alpha$) iff the (physical objective) probability of $T$'s erroneous rejection of the tested hypothesis is low (no more than $\alpha$).
	\eed 
The probability involved is presumably a physical objective probability; it represents an objective property of some process in the actual world. Consequently, a test is automatically unjustified if it turns out to lack a low probability of erroneous rejection in the actual world, irrespective of the first-person perspective of the scientist conducting the test. It does not matter whether the scientist can provide a good reason for believing or disbelieving that the test in question has a low probability of erroneous rejection. Thus, classical hypothesis testing is rendered externalist.

A note on terminology: You may have encountered variants of the above presentation that refer to Type I and Type II error probabilities. For clarity, I will largely avoid these technical terms and instead use more descriptive ones, referring directly to the probability of erroneous rejection (Type I error probability) and the probability of failing to correctly reject (Type II error probability).

Returning to the definition of significance levels, a similar externalist tone also emerges in the presentation style of Mayo \& Spanos (2011, pp. 164, 168):	
\oop 	
	{\bf Informal Definition 2.} A test is said to have a low significance level (at level $\alpha$) exactly when, if the tested hypothesis were true, $T$ would have a small probability of erroneously rejecting the tested hypothesis.
	\eed  
The truth or falsity of the counterfactual involved is, once again, an objective feature of the actual world---an external factor, independent of the first-person perspective of the scientist conducting the test.

The externalist impression might have been present for some time. When Nozick (1981) develops the tracking theory of knowledge, he notes that his two tracking conditions (in his analysis of knowledge) parallel two evaluative standards in classical hypothesis testing (1981, p. 260). In particular, the first of his tracking conditions, {\em adherence}, requires that if the hypothesis in question were true, one would believe it. This is essentially a non-probabilistic counterpart of the criterion for a low significance level, provided that Informal Definition 2 is adopted.\footnote
	{The second tracking condition, {\em sensitivity}, requires that if the hypothesis in question were false, one would not believe it. This corresponds to the criterion of a high power---a criterion to be discussed below.}
Given the parallelism Nozick draws, Fletcher \& Mayo-Wilson (2024, sec. 2) conclude that frequentist statistics is reliabilist, which might lead readers to infer that it is externalist. Otsuka (2023, sec. 3.3) explicitly claims that frequentist statistics is externalist.

%The view of frequentist statistics as externalism may reconstructed by the following master argument: 
	%{\bf Premise 1. Frequentist statistics is reliabilist, because the justification of a test in frequentist statistics depends at least in part on the significance level of a test, which is the (physical, objective) probability of erroneous rejection, measuring the (un)reliability for preventing some kind of error.   
	%{\bf Premise 2.} Reliabilism implies externalism.
	%Therefore, frequentist statistics is externalist.

However, the above two informal definitions of significance levels are actually quite misleading, as we will see when examining the textbook-standard formal definition that is employed in the practice of frequentist hypothesis testing (Section \ref{sec-formal-def}). To anticipate, the formal definition will be used to motivate and defend the following informal definition:
	\oop
	{\bf Informal Definition 3} ({\em My Preferred Choice}). A test $T$ for testing hypothesis $H_0$ has a low significance level in a context of inquiry iff, {\em for every possible world $w$ in which the background assumptions of that context are true}, if hypothesis $H_0$ is true in world $w$, then test $T$ has a low probability of (erroneously) rejecting $H_0$ in (the same) world $w$.
	\eed 
Notably, a test has a significance level only with respect to a set of background assumptions or a context of inquiry in which certain background assumptions are made---a test, therefore, does not have a significance level {\em simpliciter}. All of this is compatible with internalism. The context can be an agent's context of inquiry, and the background assumptions can be those the agent takes for granted in that context; thus, the quantification, as italicized above, can range over the worlds that {\em the agent deems (epistemically) possible from within the first-person perspective}. Under this first-person reading, a test $T$ has a low significance level exactly when the agent can deduce, from the background assumptions they take for granted, that if the tested hypothesis is true, then $T$ has a low probability of (erroneous) rejection. This concerns adducing one's background assumptions as deductive reasons for believing in a low probability of some type of error. It is entirely about evaluation and reason-giving {\em from within} one's first-person perspective. Hence internalism.

To argue for this, I will need to walk you through the formal definition of significance levels.

%\end{document}

\section{Formal Definitions Examined}\label{sec-formal-def}

My internalist interpretation of significance levels is inspired by the formal definition found in many standard textbooks, such as Casella \& Berger's {\em Statistical Inference} (2002, pp. 383-5, definitions 8.3.1, 8.3.5, and 8.3.6). Below is my slight reformulation of their definition, which removes unnecessarily technical symbols and uses underlines to highlight parts that require careful interpretation:
	\oop 
	{\bf Definition (Significance Level).} A \uline{test $T$} for testing a hypothesis $H_0$ is said to have a {\em significance level} at $\alpha$ iff $T$ satisfies the following condition with respect to the given \uline{parameter space $\Theta$}:
		\op
		\xm for every \uline{parameter value $\theta$} $\in \Theta$ that makes hypothesis $H_0$ true,
		\\ $\PP_\theta\left( T \text{ rejects } H_0 \right)$ $\le \alpha$,
		\ed 	
	where \uline{$\PP_\theta$ is the probability distribution indexed by $\theta$}.
	\eed 
\noindent To anticipate, we will see that the quantification over $\Theta$ can plausibly be understood as quantification over the worlds one deems epistemically possible, which is key to the internalist interpretation.

The underlined technical terms will be explained using an example. Imagine that we are scientists confronted with an empirical problem. There is an urn containing some marbles, and we want to test the following hypothesis:
	\begin{center}
	$H_0$: At least half of the marbles are red. 	
	\end{center}
We will stir the urn thoroughly, draw a marble, observe its color (red or non-red), replace it, and repeat the process. Once the number of observations reaches a prescribed value $n$, say $n = 4$ for concreteness, we will decide whether to reject $H_0$ (at least tentatively). In this setting, the technical terms can be explained as follows.

\subsection{Technical Term 1: Tests $T$}

An inference method for the present task---or a {\em test}---is formally a function. It can receive any data sequence of color reports of the prescribed length $n \,(=4)$, such as (\textsf{Red, Red, Non-Red, Red}), and then output a verdict: either `\textsf{Reject $H_0$}' or `\textsf{Don't}'. While there are many possible tests,\footnote
{
The total number of tests is $2^{2^n}$, which is the number of possible outputs, $2$, raised to the power of the number of possible data sequences, $2^n$.
}
the candidate pool can be narrowed down by applying certain criteria. Most notable is the criterion of a low significance level, which is widely considered the minimum qualification for {\em good} tests---or {\em justified} tests, to use a term more relevant to epistemologists.

\subsection{Technical Term 2: Parameter Values $\theta$}

Each parameter value $\theta$ denotes a possible world. For example, the world $\theta = 0.7$ is a possible world in which the proportion of red marbles in the urn is $0.7$. In this world, the tested hypothesis $H_0$, which asserts that the true proportion is at least 0.5, is true. In the world $\theta = 0.3$, the tested hypothesis is false.

\subsection{Technical Term 3: Indexed Probability Measures $\PP_\theta$}

Each parameter value $\theta$ in $\Theta$ is associated with a probability measure $\PP_\theta$, which seems to have only one sensible interpretation: $\PP_\theta$ denotes the true probability measure in world $\theta$. This measure is defined on relevant possible events, such as:
	\op 
	\im[(i)] $\PP_{\theta} \big( \,(\textsf{Red}, \textsf{Red}, \textsf{Non-Red}, \textsf{Red})\, \big)$, which denotes the probability, in world $\theta$, of obtaining data sequence $(\textsf{Red}, \textsf{Red}, \textsf{Non-Red}, \textsf{Red})$;
	
	\im[(ii)] $\PP_{\theta} \big( \text{$T$ rejects $H_0$} \big)$, which denotes the probability, in world $\theta$, that test $T$ rejects hypothesis $H_0$---that is, the probability, in world $\theta$, of obtaining a data sequence of the prescribed length $n$ that $T$ (as a function) maps to the judgment `{\sf Reject} $H_0$'.
	\ed 
The rejection probability $\PP_{\theta} \big( \text{$T$ rejects $H_0$} \big)$ is particularly important because it measures the relevant performance of a test in world $\theta$. While `performance' is the preferred term in statistics and machine learning, epistemologists can safely interpret it as `reliability'. Thus, in a world where the tested hypothesis is true, high reliability corresponds to a {\em low} probability of rejection. Conversely, in a world where the tested hypothesis is false, high reliability corresponds to a {\em high} probability of rejection.

\subsection{Technical Term 4: Parameter Space $\Theta$ (The Crux!)}

The final formal item requiring interpretation is the parameter space $\Theta$. Each element of $\Theta$ is a parameter value $\theta$ representing a possible world. So, $\Theta$ is a set of possible worlds. The exact content of $\Theta$ is a crucial topic---it marks the watershed that determines whether this frequentist statistics is externalist or internalist.

To appreciate the decisive role of $\Theta$, let's rewrite the formal definition of significance levels by incorporating the (largely uncontroversial) interpretations presented above, while leaving the exact content of $\Theta$ unspecified:
	\oop 
	{\bf Definition (Significance Level).} A test $T$ is said to achieve a  (low) {\em significance level} at $\alpha$ iff $T$'s probability of erroneous rejection is kept uniformly low across all worlds in $\Theta$ where $H_0$ is true, or put more formally:
	\begin{eqnarray*}
	\PP_{\theta} \big( \text{$T$ rejects $H_0$} \big) 
	& \le & 
	\alpha \,,
	\end{eqnarray*} 
	for every world $\theta \in \Theta$ where $H_0$ is true.
	\eed 
As a first step to narrow down the candidate pool, we set a desirable low significance level, say $5\%$, and allow only the tests that achieve this level. A second criterion is then applied to further reduce the candidate pool:\footnote
	{
	In case you are wondering how this relates to being ``powerful'': the rejection probability $\PP_{\theta} \big( \text{$T$ rejects $H_0$} \big)$ is technically called the {\em power} of $T$ at $\theta$. If $\PP_{\theta} \big( \text{$T$ rejects $H_0$} \big)$ is treated as a function of $\theta$ with $T$ held fixed, it is called $T$'s {\em power function}.
	}  
	\oop
	{\bf Definition (Uniform Maximum Power).}	A test $T$ with significance level $\alpha$ is said to be {\em uniformly most powerful}
	at level $\alpha$ iff, among all tests with the same significance level $\alpha$, the test $T$ is such that its probability of (correct) rejection is uniformly maximized across all worlds in $\Theta$ where the tested hypothesis is false, or put more formally:
	\begin{eqnarray*}
	\PP_{\theta} \big( \text{$T$ rejects $H_0$} \big) 
	& \ge & 
	\PP_{\theta} \big( \text{$T'$ rejects $H_0$} \big) \,,
	\end{eqnarray*} 
	for any alternative test $T'$ with the same significance level $\alpha$, and for any world $\theta \in \Theta$ where $H_0$ is false.
	\eed 
The above are two of the most important criteria in the Neyman-Pearson approach to frequentist hypothesis testing. These criteria share a salient feature: they are {\em standards of reliability}, as they evaluate the reliability of an inference method in each possible world across a certain range $\Theta$. But what is the exact content of $\Theta$?

If $\Theta$ is required to be the singleton containing only the {\em actual} world, then the above two criteria examine precisely the {\em actual} reliability of an inference method---a paradigm example of an external factor, rendering the Neyman-Pearson theory an externalist account. However, this externalist interpretation of $\Theta$ is quite implausible. In fact, in the numerous examples provided in Casella \& Berger's (2002) textbook, the parameter space $\Theta$ is never a singleton, let alone the singleton containing the actual world. This opens the door for internalists---a point to be elaborated in the next section.

% There seems to be a natural explanation of this.

\section{It Can Be Internalist}\label{sec-int}

Let us revisit the urn example, imagining that we are scientists testing the hypothesis that the proportion of red marbles is at least 0.5. Recall that $\theta$ represents a possible world where the proportion equals $\theta$. If we are comfortable assuming there are exactly 100 marbles in the urn, then the (epistemically) possible proportions take the form $\frac{a}{100}$. In this case, it is natural to let
	\begin{eqnarray*}
	\Theta &=& \left\{ \frac{a}{100} :  a = 0, 1, \ldots, 100 \right\} \,.
	\end{eqnarray*}
If, instead, we are only comfortable assuming that the total number of marbles in the urn lies somewhere between 10 and 100, it becomes natural to let:
	\begin{eqnarray*}
	\Theta &=& \left\{ \frac{a}{b} :  \;b = 10, 11, \dots, 100, \;a = 0, 1, \ldots, b \right\} \,.
	\end{eqnarray*}
Thus, it seems natural---or at least possible---to identify $\Theta$ with the set of possible worlds in which the {\em background assumptions} taken for granted in one's context of inquiry are true. In short, nothing in frequentist statistics prevents $\Theta$ from being the set of worlds that one deems epistemically possible within their context of inquiry---whether or not $\Theta$ contains the actual world, that is, whether or not one's background assumptions are in fact true. Under this interpretation, the evaluative criteria presented above (i.e., a low significance level and uniform maximum power) assess the reliability of an inference method in each of the worlds one deems possible within their context of inquiry. Therefore, frequentist statistics {\em can} be given an internalist interpretation.

Let's compare this internalist interpretation with an externalist one, which always sets $\Theta$ as the singleton containing only the actual world. As mentioned earlier, Casella \& Berger's (2002) textbook includes no example where $\Theta$ is a singleton---a point that aligns better with the internalist interpretation proposed above.

While the singleton-based externalist interpretation is implausible, it is not difficult to develop a second externalist interpretation. The idea is that, even if $\Theta$ is not a singleton, it can be treated as the set of worlds compatible with what one {\em knows}. This interpretation aligns with Williamson's (2000) knowledge-first philosophy. To see why this interpretation is externalist, note that knowledge implies truth. Therefore, under this knowledge-first interpretation, the parameter space $\Theta$ must include at least the actual world, requiring the evaluative criteria presented above to assess at least the actual reliability of an inference method---a paradigm example of an external factor.

I do not intend to preclude every externalist interpretation of frequentist statistics. In fact, I believe there are two kinds of justifications---internalist and externalist---each playing a distinct and valuable role in our epistemic lives, following the ideas of Mackie (1976, p. 217), Sosa (1991, p. 240), and BonJour (2005, p. 258). My emphasis is that there exists a plausible internalist interpretation.

%p. 375, Example 8.2.2 (Normal LRT)
%p. 378, Example 8.2.6 (Normal LRT with unknown variance)

% p. 409, exercise 8.41: assumption; 
% p. 409, exercise 8.42: tenable or not 

Now, let me elaborate on the internalist interpretation I have in mind. In the urn example, I stated that each parameter value $\theta$ is interpreted as a possible world in which the proportion of red marbles equals $\theta$, period. While this is correct, it is crucial to be more specific: $\theta$ should represent a world that is {\em specific enough} to ensure a unique probability measure $\PP_\theta$, the probability measure uniquely true in that world. Thus, $\theta$ is a world where the proportion of red marbles equals $\theta$ {\em and} the background assumptions hold. In the urn example, the background assumptions are:
	\oop 
	{\bf Assumption 1.} There are exactly 100 marbles (of equal size) in the urn, so that the possible proportions of red marbles form this set:
	\begin{eqnarray*}
	\Theta &=& \left\{ \frac{a}{100} :  \text{$a = 0, 1, 2, \ldots, 100$} \right\} \,.
	\end{eqnarray*}
	\eed 
	\oop 
	{\bf Assumption 2 (IID Bernoulli).} (i) Every draw is followed by replacement. (ii)  In each draw, all marbles in the urn have an equal probability of being selected. (iii) The results of all draws are probabilistically independent. 
	\eed 
The second assumption is called `IID Bernoulli' because it is short for \emph{i}ndependent and \emph{i}dentically \emph{d}istributed \emph{Bernoulli} random variables. It is taken for granted in the present context because we have agreed to always stir the urn well before drawing a marble with replacement. This second assumption is key to ensuring that each parameter value $\theta$ determines a unique probability distribution $\PP_\theta$, as commonly taught in elementary statistics (but see Appendix \ref{app-determine} for an informal, accessible presentation). In contrast, the first assumption, which restricts the possible proportions of red marbles, plays a different role: it rules out certain worlds from the parameter space $\Theta$. Consequently, $\Theta$ contains exactly the worlds in which all the background assumptions hold.

Now we are in a position to see how frequentist hypothesis testing can facilitate internalist, first-person assessments of inference methods. In frequentist statistics, an inference method is evaluated based on its reliability in each of the possible worlds across a range, formally represented by a parameter space $\Theta$. The elements of $\Theta$ represent the worlds deemed possible from one's first-person perspective---precisely the worlds in which one's background assumptions hold. This is reflected in Neyman-Pearson hypothesis testing, where the two criteria---a low significance level and uniform maximum power---are defined by quantifying over the possible worlds in $\Theta$. Thus, these criteria are defined only relative to $\Theta$ as a placeholder. A scientist making a first-person assessment must specify their own parameter space $\Theta$, delineated by the background assumptions they take for granted in their context of inquiry. In fact, frequentist standards for assessing inference methods are consistently defined relative to a parameter space $\Theta$, which serves as a placeholder to be filled in to reflect the first-person perspective from which one evaluates inference methods---this is the gist of the internalist interpretation of frequentist statistics I propose.

For clarity, let me distinguish three interconnected elements of this internalist interpretation:
\op 
\im[ ($a$)] the first-person perspective from which one assesses inference methods;
\\[-2.2em]

\im[ ($b$)] the background assumptions that one takes for granted in one's context of inquiry; 
\\[-2.2em]

\im[ ($c$)] the parameter space $\Theta$, which contains exactly the possible worlds in which the reliability of an inference method is examined.
\ed 
These three elements are closely related. Central to the internalist assessment is the first-person perspective ($a$), which may seem abstract but can be concretely characterized by articulating one's background assumptions ($b$), as demonstrated in the urn example with Assumptions 1 and 2. While the background assumptions ($b$) can be clearly expressed in plain language (suitably supported by probabilistic concepts), they can also be conveniently formalized by a parameter space ($c$), relative to which evaluative standards are defined and inference methods are assessed. Thus, the parameter space ($c$) serves as a formal representation of the first-person perspective ($a$), from which inference methods are evaluated. It is this interconnected trio, ($a$)-($c$), that enables an internalist interpretation of frequentist statistics.

%I have argued that evaluative standards in frequentist hypothesis testing have a key internalist feature: they do not concern the actual performance of an inference method, but its (varying) performance across the possible worlds in the parameter space $\Theta$, which is a formal representation of the background beliefs or assumptions that one has in one's context of inquiry. Let's be careful: I have only argued that the frequentist assessment of inference methods is internalist {\em given the background assumptions one takes for granted in one's context of inquiry}. But I have not said anything about the justificatory status of one's background assumptions. To ensure that all justifications are exclusively internalist, there still needs to be an internalist account of how one'€™s background beliefs or assumptions can be justified, in a way sensitive to one's context of inquiry. Such accounts are available in the existing literature on {\em contextualist foundationalism}; the classic text is Annis (1978), with survey articles such as Greco (@). 

The quantification over the parameter space $\Theta$ in context, which is key to enabling the internalist interpretation, is ubiquitous in frequentist statistics. While the examples above are drawn from hypothesis testing, I will provide additional examples from another inference task: point estimation (in Section \ref{sec-extension}). Before that, however, there is a more urgent matter to address.

\section{Diagnosis: Whence the Externalist Impression?}\label{sec-diagnosis}

I hope it is now clear that the Neyman-Pearson theory of hypothesis testing---and frequentist statistics more broadly---allows for an internalist interpretation. Nevertheless, it is worthwhile to step back and consider why the literature has often conveyed a strong externalist impression.

Consider a highly influential introductory textbook, Rosner's {\em Fundamentals of Biostatistics}, which, as of December 16th, 2024, has garnered more than ten thousand citations according to Google Scholar:
	\op 
	\im[$(A_1)$] The probability of a type I error is the probability of rejecting the null hypothesis when $H_0$ is true. (Rosner 2016, p. 213)

	\im[$(B_1)$] The probability of a type I error is usually denoted by $\alpha$ and is commonly referred to as the significance level of a test. (Rosner 2016, p. 214)
	\ed
This two-part presentation is not only common in introductory statistics textbooks but also adopted in some influential works in the philosophy of statistics. As Hacking writes in his {\em Logic of Statistical Inference}:
	\op 
	\im[$(A_2)$] According to this theory, there should be very little chance of mistakenly rejecting a true hypothesis. Thus, if $R$ is the rejection class, the chance of observing a result in $R$, if the hypothesis under test is true, should be as small as possible. (Hacking 1965/2016, p. 84)
	
	\im[$(B_2)$] This chance is called the size of the test; the size used to be called the significance level of the test. (Hacking 1965/2016, p. 84)
	\ed 
Similarly, in Howson \& Urbach's {\em Scientific Reasoning}:
	\op 
	\im[$(A_3)$] [I]f $H_0$ is true, the probability of a rejection ... is ... the probability of a type I error associated with the postulated rejection rule. (Howson \& Urbach 2006, p. 146) 
	
	\im [$(B_3)$] This probability is called, as before, the significance level of the test. (Howson \& Urbach 2006, p. 146) 
	\ed 
The above presentations all consist of two parts, an $A$-part and a $B$-part. The problem arises when too much focus is placed on the $B$-parts---i.e., $(B_1)$, $(B_2)$, and $(B_3)$---which tend to create the impression that the significance level of a test is defined as {\em the} probability of a certain event. This is easily mistaken to mean {\em the} probability of a certain event in {\em the} actual world. Since probabilities in the actual world are external factors, independent of one's background assumptions or first-person perspective, this creates the externalist impression.

To guard against this externalist implication, recall that a low significance level actually conveys something else: it means that, in every world $\theta \in \Theta$ where $H_0$ is true, the Type I error probability (i.e., the probability of rejecting $H_0$) is low. In other words, a significance level imposes an upper bound on the Type I error probabilities across the worlds in $\Theta$ (the least upper bound is known as the {\em size}). The key, once again, lies in the quantification over the parameter space $\Theta$, which formally represents one's background assumptions or knowledge---the former being internalist and the latter externalist. Thus, the internalist interpretation is not automatically precluded.

Rosner's influential textbook takes steps to prevent misunderstanding. In the context where he introduces the two-part definition---namely, chapter 7 of Rosner (2016)---the focus is on testing a {\em point} hypothesis $H_0$, such as ``the proportion of red balls in the urn is {\em exactly} $50\%$'' or ``the true mean of an unknown normal distribution is {\em exactly} $0$.'' These are called point hypotheses because there is exactly one parameter value $\theta_0 \in \Theta$ that makes the tested hypothesis $H_0$ true, with $\theta_0$ representing a single point in the parameter space. In such cases, we can unambiguously refer to {\em the} Type I error probability, which is a probability in {\em the} world $\theta_0$, without mistaking it for a probability in {\em the} actual world. Instead, it refers to the probability of rejection in the unique world $\theta_0 \in \Theta$ that makes $H_0$ true.

When statisticians move on to testing a {\em composite} hypothesis, which is true in {\em multiple} worlds in $\Theta$, the definition of significance levels must explicitly involve quantification over these possibilities. This is illustrated in another elementary textbook, authored by Ross:
	\begin{quote}
	The classical way of accomplishing [the desideratum expressed above] is to specify a small value $\alpha$ and then require that the test have the property that {\em whenever} $H_0$ is true, its probability of being rejected is less than or equal to $\alpha$. The value $\alpha$ [is] called the level of significance of the test. (Ross 2010, p. 391, emphasis mine) 	
	\end{quote}
The phrase `whenever' is well-chosen, better than the use of `when' as in $(A_1)$. Indeed, `whenever' more clearly signals that a quantification `for all' is involved. This is further clarified by the example Ross provides on the same page: testing the hypothesis that the mean nicotine level of certain cigarettes is greater than or equal to 1.5 units---a composite hypothesis. In this case, the significance level must be defined by quantification over all $\theta \ge 1.5$.

Instead of `whenever,' Mayo uses an indicative conditional `if' in her influential defense of frequentist statistics (Mayo 1996, p. 180), which avoids any unintended temporal connotations. More importantly, she carefully reminds us, on the same page, of the ``underlying assumptions or background conditions,'' which reinforces the internalist reading and appropriately restricts the range of possible worlds that `if' quantifies over.

But even Mayo occasionally creates an externalist impression. In her collaboration with Spanos, she continues to use the conditional `if', but this time in a counterfactual form (Mayo \& Spanos 2011, pp. 164, 168). Setting aside the nuanced differences between the Mayo-Spanos view and the Neyman-Pearson view, the counterfactual formulation reads as follows:
	\oop 
	{\bf A Counterfactual-Based Informal Definition of Significance Levels.} A test has a low significance level just in case, if $H_0$ were true, the test would have a low probability of rejecting $H_0$. 
	\eed  
Taken literally, this presents an externalist account. Whether the counterfactual on the right-hand side is true or false is an external factor---it involves something independent of one's background beliefs or assumptions. This externalist interpretation becomes even more pronounced if the reader adopts the similarity semantics of counterfactuals (Stalnaker 1968), which is well-known in the philosophical community. In this case, the counterfactual formulation becomes the following:
	\oop 
	{\bf A Similarity-Based, Informal Definition of Significance Levels.} A test has a low significance level just in case, in the closest-to-actuality world in which $H_0$ is true, the test has a low probability of rejecting $H_0$. 
	\eed
Recall that, in the urn example, $H_0$ is true in the worlds where the proportion of red marbles is at least 50\%. So, if the actual world is $\theta = \frac{70}{100}$, the closest world that makes $H_0$ true is the same world, $\frac{70}{100}$, and the probability referred to is a probability in that world. However, if the actual world is $\theta = \frac{30}{100}$ instead, the closest world that makes $H_0$ true seems to be $\frac{50}{100}$, and the probability referred to becomes a probability in that world, $\frac{50}{100}$. Thus, the referent of `probability of rejecting $H_0$' depends on which world is actual, regardless of one's background assumptions. This is clearly an externalist interpretation.

Mayo \& Spanos (2011) likely do not intend their account to be committed to externalism. In fact, they dedicate two pages to emphasizing the importance of {\em background knowledge} (p. 159), {\em background information} (p. 159), and {\em background opinions} (p. 160), leaving open both externalist and internalist interpretations. Unfortunately, this discussion appears four pages before they introduce the counterfactual formulation (pp. 164, 168), which, when read literally, leans toward an externalist interpretation.

Informal definitions are often valuable---especially when we want to focus on philosophical ideas without the distraction of formal symbols. So, when an explicit reference to a parameter space $\Theta$ seems too formal or technical, I recommend that the criterion for a low significance level be informally defined as follows:
	\oop 
	{\bf A Better Informal Definition of Significance Levels.} A test $T$ has a (low) significance level at $\alpha$ iff $T$ is guaranteed, under the background assumptions, that whenever $H_0$ is true, $T$ has a no-more-than-$\alpha$ probability of (erroneously) rejecting $H_0$.  
	\eed 
Similarly for the criterion of uniform maximum power at a significance level:
	\oop  
	{\bf A Better Informal Definition of UMP Tests.} A test $T$ is uniformly most powerful at a (low) significance level $\alpha$ iff, first, $T$ has a significance level at $\alpha$ and, second, $T$ is guaranteed, under the background assumptions, that whenever $H_0$ is false, $T$ has the maximum probability of (correctly) rejecting $H_0$ subject to the constraint of a significance level at $\alpha$.
	\eed 

I hope this dispels the misconception that frequentist statistics necessarily leads to externalism. The quantification over the parameter space $\Theta$---viewed as a set of possible worlds---enables the development of an internalist interpretation. Omitting the domain of quantification $\Theta$ distorts the actual statistical practice, as seen in the counterfactual formulation and in formulations that identify a significance level with a single probability. Even if one wishes to allow for an externalist stance---whether as a thoroughgoing externalist or a compatibilist who permits both internalist and externalist justifications to play their respective roles---it is still better to stay as close to the actual statistical practice as possible by retaining the domain of quantification $\Theta$ and using it to represent one's background knowledge, information, assumptions, beliefs, or whatnot.

	%\oop A test has a low significance level (at $\alpha$) iff, whenever $H_0$ is true, that test has a low probability of rejecting $H_0$.\eed 

%\end{document}

\section{It's (Unconventionally) Reliabilist}\label{sec-relia}

Frequentist statistics can be not only internalist but also {\em simultaneously} reliabilist---in a broader sense of reliabilism that does not imply externalism. In fact, there are two unconventional (but closely related) senses in which frequentist statistics is reliabilist. Let me explain.

\subsection{Reliabilism in a Broader Sense}

Frequentist statistics is reliabilist in at least this sense:
	\oop 
	{\bf Frequentist Statistics as Reliabilism 1.} In frequentist statistics, inference methods are always assessed by {\em standards of reliability}---standards that examine the \uline{relevant reliability} of an inference method in each of the possible worlds across one or another \uline{range $\Theta$}. 
	\eed 

The underlines highlight two key concepts. The first one---the relevant reliability---is sensitive to one's context of inquiry. When the inference task is hypothesis testing, the reliability of an inference method is often defined in terms of the probability of (correct or erroneous) rejection, as seen above. For another example, when the inference task is interval estimation, where we aim to produce an interval as an estimate of an unknown quantity, the reliability of an inference method is defined as the probability of producing a (short) interval that covers the true value of the estimated quantity. A similar example arises in point estimation, where we aim to produce a point estimate; the reliability of an inference method can be defined as the probability of producing a point close to the true value, but it is also commonly defined by the so-called {\em mean squared error}.\footnote
{Here is an informal explanation of the concept of mean squared error. Let $M$ be an estimator. When $M$ receives a data sequence, it produces an estimate and incurs an error, which is the difference between the estimate and the true (but unknown) value of the estimated quantity. There are many possible errors: given one data sequence, $M$'s error might be small; given another, it might be large. The weighted average of the squared possible errors, with weights being the probabilities over the possible data sequences, is the mean squared error of estimator $M$.}
There are additional inference tasks in statistics, such as model selection, regression, and classification, and the point I make still applies; see Lin (2024$b$) for examples. In general, when switching to a new inference task, we may need to redefine the conception of reliability in use---to select the relevant reliability.

The second underline indicates a set of possible worlds, $\Theta$, which is also sensitive to one's context of inquiry and allows for {\em both} internalist and externalist interpretations, as seen above. This is important for clarifying the logical relationship between externalism and reliabilism. The traditional wisdom that reliabilism implies externalism holds true when we limit ourselves to the {\em conventional} sense of reliabilism, according to which the factors of justification must include at least the {\em actual} reliability of the relevant inference method or belief-producing procedure. However, frequentist statistics is reliabilist in a broader sense: the evaluative standards in use are all defined to examine the reliability of an inference method in certain worlds---the worlds in $\Theta$. When $\Theta$ is set to be the singleton containing just the actual world, frequentist statistics specializes into a reliabilist theory in the conventional sense, somewhat akin to Goodman's (1972) process reliabilism. When $\Theta$ is identified with the set of worlds compatible with what one knows, frequentist statistics aligns with Williamson's (2000) knowledge-first epistemology, retaining a conventional reliabilist interpretation. But when $\Theta$ is identified with the set of worlds in which one's background assumptions hold, frequentist statistics becomes reliabilist in an unconventional sense.

Therefore, reliabilism in the broader sense is compatible with internalism---with a distinctive example taken from the scientific practice: frequentist statistics. 

\subsection{Achievabilist Norms in Reliabilism}\label{sec-achieva}

Frequentist statistics is reliabilist in an additional sense: the choice of the operative standard for assessing an inference method is context-sensitive; it is set to be the {\em highest achievable} standard of reliability---achievable within the context of the problem at hand. This embodies a serious pursuit of reliability. Let me walk you through some examples.

Recall the urn case discussed above, where the tested hypothesis extends to one side of the real line:
	\op 
	\im[] $H_0:$ ``The proportion of red marbles is at least $50\%$.''
	\ed 
In this case, there exists a test that achieves the high standard of uniform maximum power at a low significance level (thanks to an extension of the Neyman-Pearson lemma, known as the Karlin-Rubin theorem).\footnote{For a textbook presentation of the Neyman-Pearson lemma, see Casella \& Berger (2002, p. 388). For the Karlin-Rubin theorem, see Casella \& Berger (2002, pp. 391-392).} We should then aim for this high standard. However, this standard might become too high to achieve when we switch to other problem contexts. For example, suppose we are now testing a hypothesis that is restricted on both sides of the real line, such as:
	\op 
	\im[] $H_0:$ ``The proportion of red marbles is equal to $50\%$.''
	\ed 
or 
	\op 
	\im[] $H_0:$ ``The proportion of red marbles is in $[45\%, 55\%]$.''
	\ed 
In such a ``two-sided'' problem, it is provable that no test achieves the standard of uniform maximum power at any given significance level, let alone at a low significance level (Casella \& Berger 2002, pp. 392-393, example 8.3.19).

A possible reaction is to settle for a single, lower standard for all problems of hypothesis testing. However, this is {\em not} the reaction recommended by frequentist statisticians. In the ``one-sided'' problem, a high standard is achievable, so anyone tackling that problem is required to strive for that high standard. One may settle for a lower standard only when mathematical necessity dictates it---only when no test can achieve the high standard in the given problem context. The ``two-sided'' problem is one such example. It is only in such cases that frequentist statisticians resort to a lower standard.

%The guiding principle is to strive for the {\em highest achievable} standard in each problem undertaken---an idea that was already present in the minds of Neyman \& Pearson (1936), two of the founding parents of frequentist hypothesis testing.

It remains to find a sensible lower standard that is achievable in the ``two-sided'' problem. Let's begin by revisiting the higher standard---uniform maximum power at a low significance level---and offering a revealing reformulation:
	\op
	\im First, narrow down the candidate pool by ruling out the tests that fail the criterion of a low significance level.
	\im Then, require that the probability of correct rejection be maximized at each world in $\Theta$ where $H_0$ is false---maximized among the candidates remaining from the previous step. 
	\ed 
This approach narrows down the candidate pool one step at a time. The second step, maximization, can be quite demanding, and at times, it may be too demanding to achieve. The larger the candidate pool left from the previous step, the more demanding the maximization becomes, as it involves maximization over all the remaining candidates. At this point, it's not hard to think of a lower standard: postpone the maximization step until we have a smaller candidate pool. This idea has a textbook-standard implementation (Casella \& Berger 2002: p. 393, example 8.3.20):
%This idea, first introduced by Neyman \& Pearson (1936), is now textbook-standard (Casella \& Berger 2002: p. 393, example 8.3.20):
	\op
	\im First, narrow down the candidate pool by using the criterion of a low significance level, that is, by requiring that the probability of (erroneous) rejection be low (at most $\alpha$) whenever the $H_0$ is true.
	\im Second, narrow down the candidate pool further by requiring that the probability of (correct) rejection be at least not too low (e.g., at least $\alpha$) whenever $H_0$ is false.
	\im Then, require that the probability of correct rejection be maximized at each world in $\Theta$ where $H_0$ is false---maximized among the candidates remaining from the previous step. 
	\ed 
Note the additional filter applied before the final maximization step. This extra filter (the second step) rules out some candidates and retains only those known as the {\em unbiased} tests (Casella \& Berger 2002: p. 387, definition 8.3.9). Therefore, in the final step, the probability of correct rejection is maximized over a smaller candidate pool.

We now have a hierarchy of standards of reliability, which are defined with respect to two contextual factors, (i) a hypothesis $H_0$ to be tested, and (ii) a parameter space $\Theta$ representing one's background assumptions:
$$\begin{array}{l}
	\textit{Uniform Maximum Power}
\\
	\textit{Among the Tests at a Low Significance Level $\alpha$}
\\
	\quad\quad\quad\quad\quad |
\\
	\textit{Uniform Maximum Power}
\\
	\textit{Among the Unbiased Tests at a Low Significance Level $\alpha$}
\\
	\quad\quad\quad\quad\quad |
\\
	\textit{A Low Significance Level $\alpha$ (Minimum Qualification)}
\end{array}$$
Formal definitions are provided in Appendix \ref{app-formal} for reference. 

Frequentist hypothesis testing does not apply a single standard of reliability across all problem contexts. In practice, the operative standard is ideally the highest achievable. Therefore, I propose that frequentist hypothesis testing operate under the following norm, even though statisticians seem to have never stated this norm explicitly:
	\oop 
	{\bf Achievabilist Reliabilism in Hypothesis Testing.} For every problem context $C$ that specifies a hypothesis $H_0$ slated for testing and a parameter space $\Theta$ representing one's background assumptions or knowledge, a test is justified in context $C$ only if it meets the highest standard of reliability that is achievable with respect to $H_0$ and $\Theta$---pending a specification of the correct hierarchy of standards. 
	\eed 
According to this norm, the operative standard of reliability is sensitive to what is achievable in the specific context at hand---hence, it is an {\em achievabilist} norm. This leads to an achievabilist version of reliabilism.

\subsection{Extensions}\label{sec-extension}

What I've just outlined applies not only to hypothesis testing but also extends to other inference tasks studied in frequentist statistics, such as point estimation, interval estimation, model selection, (nonparametric) regression, and classification. Let me provide an example from point estimation.\footnote{For examples in model selection and classification, see Lin (2024$b$).}

In the standard textbook by Lehmann \& Casella (1998), {\em Theory of Point Estimation}, various criteria for assessing point estimators are defined. Let me mention a few examples:
	\op 
	\im There is the minimum qualification known as {\em admissibility}, which means freedom from having the relevant reliability being dominated by an alternative estimator across the worlds in the given parameter space $\Theta$, where the relevant reliability is conventionally identified with the mean squared error (Lehmann \& Casella 1998, p. 48).
	
	%\im A higher standard adds {\em minimax} (which requires maximizing the relevant reliability in the worst-case world). 6 p. 306, definition 1.1
	
	\im We obtain a higher standard by conjoining admissibility with {\em unbiasedness}, which means, very roughly, that the expected overestimation matches the expected underestimation across the worlds in $\Theta$ (Lehmann \& Casella 1998, p. 83, definition 1.1).
	
	\im An even higher standard adds to admissibility and unbiasedness a property known as {\em UMVU}, short for {\em u}niformly {\em m}inimizing the {\em v}ariance among the {\em u}nbiased estimators. This means that the relevant reliability is uniformly maximized across the worlds in $\Theta$ among the unbiased estimators (Lehmann \& Casella 1998, p. 85, definition 1.6).
	\ed 
Caveat: While admissibility is widely regarded as a minimum qualification in point estimation, unbiasedness remains somewhat controversial, despite its extensive coverage in almost all standard textbooks.\footnote{See Lehmann \& Casella (1998, pp. 5, 157-158) for a controversy surrounding unbiasedness and a possible alternative (known as {\em median-unbiasedness}). Also see Jaynes (2003, sec. 17.3) for a discussion.} Indeed, determining the correct hierarchy of standards of reliability is an issue open to exploration and debate. Even so, the pursuit of the highest achievable standard still seems to lie at the heart of frequentist statisticians.

I hereby propose the following norm to capture an important aspect of the practice of frequentist statisticians in general:\footnote{
The first achievabilist norm, stated at a high level of generality, is due to Lin (2022), who develops a counterpart of the present statement in the context of a non-stochastic theory of scientific inference, formal learning theory. Lin (forthcoming) extends the achievabilist norm to cover both the stochastic and non-stochastic settings simultaneously. A note on terminology: The achievabilist norm is called {\em the core thesis of learning-theoretic epistemology} in Lin (2022, p. 284). This name is aptly descriptive, as the achievabilist norm is indeed central to learning theory, including both formal learning theory in philosophy and statistical learning in machine learning. However, this name has a downside: it might create the false impression that the spirit of striving for the highest achievable standard is unique to learning theory. In fact, this spirit is also core to frequentist statistics, as I have argued here. This is why I adopt the more neutral term `achievabilism', following the usage in Lin (forthcoming).
}
	\oop 
	{\bf Frequentist Statistics as Reliabilism 2 (Achievabilist Reliabilism).} 
	For any problem context $C$, an inference method is justified in $C$ only if it meets the highest standard of reliability that is achievable in context $C$---pending a specification of the correct hierarchy of standards. 
	\eed 
Caveat: This statement is only meant as a first approximation. Complications arise if the correct hierarchy is not a linear order but only a {\em partial} order (allowing for two incommensurable standards, neither of which is higher than the other, nor are they equal), or if there is no {\em uniquely} highest achievable standard (possibly because there are many, or none), or if there is no such thing as {\em the} correct hierarchy. In any of these cases, the statement of the norm would need to be revised accordingly. Nonetheless, my point remains: Frequentist statisticians do not merely use standards of reliability to assess inference methods; they also strive for {\em a} (if not {\em the}) highest achievable standard.

As far as I know, no statisticians explicitly state this norm at such a general level. However, their textbooks are filled with definitions of various standards of reliability, often indicating which ones are higher or lower, along with numerous examples of problem contexts where one standard or another is shown to be achievable or unachievable, as we have seen above. Therefore, the norm stated above does seem to capture an important aspect of their practice.

Frequentist statistics is therefore reliabilist not only in the sense of employing standards of reliability, but also in striving for the highest achievable standard of reliability in every context of inquiry.

\section{Closing}

Perhaps it is not difficult to stretch the meaning of `reliabilism' merely to make it fail to imply externalism. However, that is not what I have done here. Instead, I have broadened related concepts and defended the thesis that reliabilism does not imply externalism for an important reason: to accommodate a natural and plausible interpretation of frequentist statistics.

Much more remains to be done to develop this internalist interpretation. First, there is the task of explaining how background assumptions may be justified within one's context of inquiry, possibly following Annis (1978), who, like me, also advocates for the context-sensitive nature of justified beliefs. Second, while emphasis has been placed on the first-person perspective for assessing inference methods, this perspective can, in principle, be extended to the first-person {\em plural}, allowing the parameter space $\Theta$ to represent the assumptions shared by the members of a community---the common ground of that community. However, the details must be left for future work.

%%% connect to Longino?

%%% mention naturalism?
\section*{Acknowledgements}

I am grateful to the participants of the Philosophy of Science and Epistemology Conference at the Hong Kong University of Science and Technology (June 27-29, 2024) for their valuable comments and questions. I am especially indebted to Jun Otsuka, Conor Mayo-Wilson, I-Sen Chen, Yafeng Shan, and an anonymous referee for their helpful comments. Special thanks go to Konstantin Genin and Kevin Kelly; my conversations with them made me realize the value of arguing for an internalist interpretation of frequentist statistics.

\appendix

\section{Appendix}

\subsection{How $\PP_\theta$ Is Determined}\label{app-determine}

Although not required for the purposes of this paper, it is helpful to have a concrete picture of how $\PP_\theta$ is determined for each $\theta \in \Theta$ without delving into too many technical details. 

In the world where the proportion of the red balls is $\theta = 0.7$, the probability of obtaining a red marble in each draw is equal to $0.7$, by clauses (i) and (ii) of IID Bernoulli (as presented in Section \ref{sec-int}):
	$$
	\begin{array}{lll}
	\PP_{0.7} \big( \textsf{Red} \big) &=& 0.7 \,,\\
	\PP_{0.7} \big( \textsf{Non-Red} \big) &=& 1 - 0.7 \,.
	\end{array}
	$$
For simplicity, let the prescribed sample size be $n = 4$. Then the probability of obtaining a data sequence, say $(\textsf{Red}, \textsf{Red}, \textsf{Non-Red}, \textsf{Red})$, can be decomposed according to clause (iii) of IID Bernoulli:
	\begin{eqnarray*}
	&& \PP_{0.7} \big((\textsf{Red}, \textsf{Red}, \textsf{Non-Red}, \textsf{Red}) \big)
\\
	&=& \PP_{0.7} \big( \textsf{Red} \big) 
	\cdot \PP_{0.7} \big( \textsf{Red} \big) 
	\cdot \PP_{0.7} \big( \textsf{Non-Red} \big)
	\cdot \PP_{0.7} \big( \textsf{Red} \big)
	\end{eqnarray*}
Combining the above results, we have:
	\begin{eqnarray*}
	&& \PP_{0.7} \big((\textsf{Red}, \textsf{Red}, \textsf{Non-Red}, \textsf{Red}) \big)
\\
	&=& 0.7 
	\cdot 0.7 
	\cdot (1 - 0.7)
	\cdot 0.7
	\end{eqnarray*}
This calculation procedure generalizes quite straightforwardly, which determines for each $\theta$ a unique probability distribution $\PP_\theta$ as a function that assigns real numbers to the $2^n$ data sequences.

Then, the value of $\PP_{\theta} \big( \text{$T$ rejects $H_0$} \big)$ can be defined and computed using the following procedure:
	\op
	\im {\em Step 1:} Start with any given test $T$ and any given world $\theta$, which is associated with a unique probability distribution $\PP_{\theta}$ that assigns probabilities to the $2^n$ data sequences. 
	
	\im {\em Step 2:} Mark every data sequence that, if received, would prompt $T$ to output `\textsf{Reject $H_0$}'. 
	
	\im{\em Step 3:} Find the probability that $\PP_{\theta}$ assigns to each of those marked data sequences. 
	
	\im {\em Step 4:} Sum these probabilities and return the result as the value of $\PP_{\theta} \big( \text{$T$ rejects $H_0$} \big)$. 
	\ed 
Thus, the probabilities involved in the urn example are all defined with respect to each parameter value $\theta$.

\subsection{Some Formal Definitions}\label{app-formal}

This paper examines some evaluative criteria in frequentist hypothesis testing. Their formal definitions are provided here for reference, with minimal interpretation to aid readability, leaving the crucial term $\Theta$ uninterpreted.

Start with the minimum qualification, a low significance level, whose formal definition has already been provided in the main text but listed here for completeness: 
	\oop 
	{\bf Definition (Significance Level).} A test $T$ is said to achieve a (low) significance level at $\alpha$ iff $T$'s probability of erroneous rejection is uniformly low in this sense:
	$$\PP_{\theta} \big( \text{$T$ rejects $H_0$} \big) \;\le\; \alpha \,,$$
	for any world $\theta \in \Theta$ where $H_0$ is true.
	\eed  
	%$$\PP_{\theta} \big( \text{$T$ rejects $H_0$} \big) \;\ge\; \alpha$$ for any world $\theta \in \Theta$ where $H_0$ is false.
The following is another evaluative criterion, which was only informally sketched in the main text: 
	\oop 
	{\bf Definition (Unbiasedness).} A test $T$ is said to be {\em unbiased} iff $T$'s probability of correct rejection is uniformly not too low in this sense:
	$$\PP_{\theta} \big( \text{$T$ rejects $H_0$} \big) \;\ge\; \PP_{\theta'} \big( \text{$T$ rejects $H_0$} \big) \,,$$
	for any world $\theta \in \Theta$ where $H_0$ is false and for any world $\theta' \in \Theta$ where $H_0$ is true. 
	\eed 
There is also a schema that, while not corresponding to any single criterion, is useful for constructing new criteria from old ones. Suppose we already have some criteria that narrow down the candidate pool to a class $\mathscr{C}$. We can then define an additional criterion as follows:
	\oop 
	{\bf Definition (Uniform Maximum Power in a Class).} A test $T$ for hypothesis testing is said to be {\em uniformly most powerful} in a class $\mathscr{C}$ of tests iff, first, $T$ belongs to class $\mathscr{C}$ and, second, $T$'s probability of (correct) rejection is uniformly maximized in this sense:
	\begin{eqnarray*}
	\PP_{\theta} \big( \text{$T$ rejects $H_0$} \big) 
	& \ge & 
	\PP_{\theta} \big( \text{$T'$ rejects $H_0$} \big) \,,
	\end{eqnarray*} 
		for any alternative test $T' \in \mathscr{C}$ and for any world $\theta \in \Theta$ where $H_0$ is false.
	\eed 
Treat the class $\mathscr{C}$ in the above as a placeholder. Once we replace $\mathscr{C}$ by a candidate pool of tests delineated by some existing criteria, we can narrow down the candidate pool further by picking out those that are uniformly most powerful in $\mathscr{C}$. When $\mathscr{C}$ is set to be the class of the tests at significance level $\alpha$, we obtain the highest standard in the hierarchy discussed in Section \ref{sec-achieva}; the second highest is obtained by letting $\mathscr{C}$ be the class of the unbiased tests at significance level $\alpha$.

Reminder: all these standards are defined with respect to at least two contextual factors: a hypothesis $H_0$ slated for testing, and a parameter space $\Theta$, whose possible interpretations are crucial to the discussion of internalism vs. externalism in statistics.

%Wanna talk about consistency here?
% Fisher on hypothesis testing
%Later Pearson on hypothesis testing?

%\end{document}

\section*{Bibliography}

\begin{description}
\im Annis, D. B. (1978). A contextualist theory of epistemic justification. {\em American Philosophical Quarterly, 15}(3), 213-219.

\im Bonjour, L. (2005). Internalism and externalism. In P. K. Moser (Ed.), \textit{The Oxford handbook of epistemology} (pp. 234-263). Oxford University Press.

\im Casella, G., \& Berger, R. (2002). \textit{Statistical inference} (2nd ed.). Duxbury Press.

\im Fletcher, S. C., \& Mayo-Wilson, C. (2024). Evidence in classical statistics. In M. Lasonen-Aarnio \& C. Littlejohn (Eds.), \textit{The Routledge handbook of the philosophy of evidence} (pp. 515-527). Routledge.

\im Goldman, A. I. (1979). What is justified belief?. In G. S. Pappas (Ed.), \textit{Justification and knowledge: New studies in epistemology} (pp. 1-23). Springer Netherlands.

\im Hacking, I. (1965/2016). \textit{Logic of statistical inference}. Cambridge University Press.

\im Howson, C., \& Urbach, P. (2006). \textit{Scientific reasoning: The Bayesian approach} (3rd ed.). Open Court Publishing.

\im Jaynes, E. T. (2003). \textit{Probability theory: The logic of science}. Cambridge university press.

\im Lehmann, E. L., \& Casella, G. (2006). \textit{Theory of point estimation}. Springer Science \& Business Media.

\im Lin, H. (2022). Modes of convergence to the truth: Steps toward a better epistemology of induction. \textit{The Review of Symbolic Logic, 15}(2), 277-310.

\im Lin, H. (2024$a$). To be a frequentist or Bayesian? Five positions in a spectrum. \textit{Harvard Data Science Review}, 6(3), doi: 10.1162/99608f92.9a53b923

\im Lin, H. (2024$b$). Internalist reliabilism in statistics and machine learning: Thoughts on Jun Otsuka's thinking about statistics. {\em The Asian Journal of Philosophy 3}(2), 1-11.

\im Lin, H. (forthcoming). Convergence to the truth. In K. Sylvan, E. Sosa, J. Dancy, \& M. Steup (Eds.), \textit{The Blackwell companion to epistemology} (3rd ed.). Wiley Blackwell.

\im Mackie, J. L. (1976). \textit{Problems from Locke}. Oxford University Press.

\im Mayo, D. G. (1996). \textit{Error and the growth of experimental knowledge}. University of Chicago Press.

\im Mayo, D. G., \& Spanos, A. (2011). Error statistics. In P. S. Bandyopadhyay \& M. R. Forster (Eds.), \textit{Philosophy of statistics} (pp. 153-198). North-Holland.

%\im Neyman, J., \& Pearson, E. S. (1936). Contributions to the theory of testing statistical hypotheses. {\em Statistical Research Memoirs, 1}, 1-37.

\im Neyman, J. (1955). The problem of inductive inference. \textit{Communications on Pure and Applied Mathematics, 8}(1), 13-45.

\im Nozick, R. (1981). \textit{Philosophical explanations}. Cambridge University Press.

\im Otsuka, J. (2023). \textit{Thinking about statistics: The philosophical foundations}. Routledge.

\im Popper, K. R. (1959). The propensity interpretation of probability. \textit{The British Journal for the Philosophy of Science, 10}(37), 25-42.

%\im Quine, W. V. O. (1969). Epistemology naturalized. In \textit{Ontological relativity and other essays} (pp. 69-90). Columbia University Press.

\im Ross, S. M. (2010). \textit{Introductory statistics} (3rd ed.). Elsevier.

\im Rosner, B. A. (2016). \textit{Fundamentals of biostatistics} (8th ed.). Cengage Learning.

\im Salmon, W. C. (2017). {\em The foundations of scientific inference: The 50th anniversary edition}. University of Pittsburgh Press.

\im Sober, E. (2000). \textit{Philosophy of biology}, 2nd Edition. Westview Press.

\im Sosa, E. (1991). \textit{Knowledge in perspective}. Cambridge University Press.

\im Stalnaker, R. C. (1968). A Theory of Conditionals. In Harper, W. L., Pearce, G. A., \& Stalnaker, R. C. (eds.) {\em Ifs: Conditionals, belief, decision, chance and time}. Springer Netherlands: 41-55.

\im Steup, M. (2004). Internalist reliabilism. \textit{Philosophical Issues, 14}, 403-425.

\im Williamson, T. (2000). \textit{Knowledge and its limits}. Oxford University Press.
\end{description}

\end{document}